# Trajectory Tracking Control of the Bionic Joint Actuated by Pneumatic Artificial Muscle Based on Robust Modeling

Yang Wang, Qiang Zhang, and Xiao-hui Xiao*

College of Power and Mechanical Engineering, Wuhan University, Wuhan 430072, China

**Abstract:** The high nonlinearity is a great challenge for the bionic system actuated by pneumatic artificial muscle (PAM). In this paper, a practical and effective method is proposed to realize a trajectory tracking of an unsymmetrical bionic joint actuated by a single PAM based on robust control theory. By applying the robust modeling method, the nonlinear model between the driving pressure input of PAM and the angular position output of the joint is represented by a linear time-invariant system with parametric perturbations. And, a cascaded controller, constructed with an outer $\mathcal{H}\infty$ controller for the position tracking and an inner controller for the driving pressure tracking, is developed, where the $\mathcal{H}\infty$ controller is synthesized by loop-shaping design procedure (LSDP) and the pressure controller based on feedback linearization.

## 1 Introduction

For the friendly interaction with human living environment, humanoid robot is always attractive for researchers of robotics. To achieve good performance and adaptability in various fields, the improvement on efficient of bionic joint is imperative. For the inherent advantages, such as high power-to-weight ratio, compliance and non-pollution, PAM has been used as a novel actuator for bionic system. However, the nonlinearity of the model between the force output, contraction displacement, and inner pressure of PAM is a great challenge for controller design to realize a perfect trajectory tracking [1].

To deal with the nonlinearity of PAM system, nonlinear control strategies are applied such as feedback linearization [2], backstepping control[3], adaptive control based on neural networks[4], and sliding mode control[5-6].

Besides these nonlinear control strategies, many linear and mixed strategies are also investigated such as predictive robust control[7] and $\mathcal{H}_\infty$ control[8].

Although the tracking accuracy is relatively low, linear control strategies are easier to implement and can adopt more useful skills in frequency domain. Therefore, in this article, a practical and effective method is proposed to reduce the effect of nonlinearity of a bionic system with PAM actuator and realize a trajectory tracking based on robust control theory.

In this paper, the system model between the actual voltage input and joint motion output is divided into two cascaded subsystems, an inner pneumatic subsystem and an outer mechanical subsystem. The former model is analytic and the later is modelled by robust identification method. Accordingly, a cascaded controller is developed, where the inner controller is based on feedback linearization for tracking the driving pressure of PAM and the outer $\mathcal{H}\infty$ controller synthesized by LSDP method for suppressing the effect of nonlinearity of mechanical motion[9].

The layout of this paper is shown as follows: Section 2 illustrates and models the pneumatic circuit and dynamics systems. Section 3 designs an $\mathcal{H}\infty$ controller cascaded with a nonlinear controller of the PAM actuator system for trajectory tracking control. Section 4 implements and analyses the tracking experiment of the controlled platform. Thereby, the trajectory tracking of a PAM system is realized with good performance.

## 2 Physiological structure of PAM platform

In this paper, a bionic elbow joint is built to simulate the back-and-forth movement of forearm in the pitching plan as shown in Fig 1. The structure of PAM commonly used is depicted in Fig 2[10]. The main part of it is the cylindrical membrane, fixed between two flanges, which is of isotropic flexibility and made by textile-fiber. While the muscle inflated by compressed air, its contracts in length direction, expands in radial and output tension force. The mechanical platform is based on the 'slider-crank' structure which can realize the transmission between rectilinear motion/force

Foundation item: Project(51175383) supported by the National Natural Science Foundation of China.
Received date:
Corresponding author: Xiao-hui Xiao, professor, PhD; Tel: 13720360269; E-mail: xhxiao@whu.edu.cn

and rotatory motion/torque as human elbow musculo-skeletal system with compact size, as shown in Fig 3[11]. The pendulum of it represents the forearm as the load of the joint. The parameters of mechanical platform are listed in Table 1.

For the compliance of PAM and unsymmetrical structure, the pair of antagonistic forces for driving the joint is the pulling force of PAM and the gravity on the pendulum. Thus the theoretical workspace is from horizontal position to vertical position.

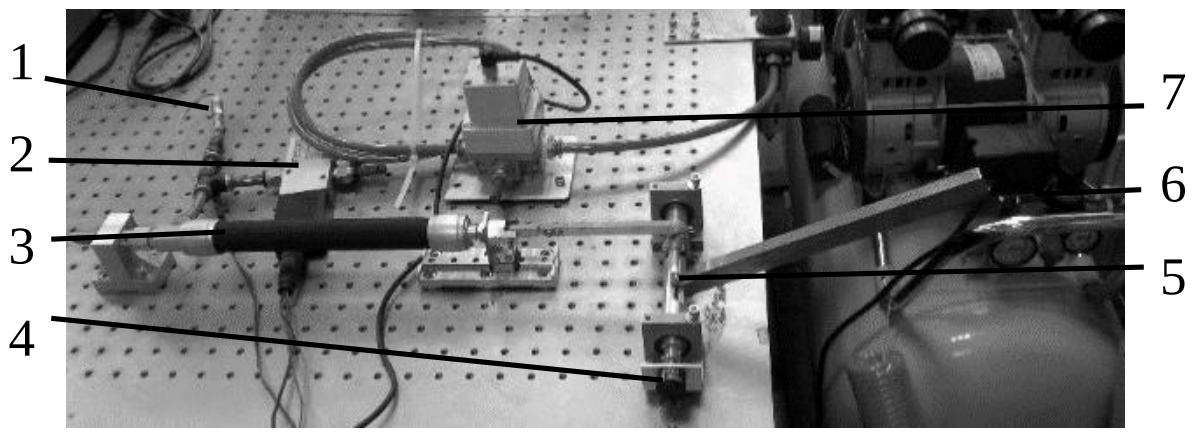

**Fig.1** Experiment platform
( 1 Air pressure transducer; 2 Electronic proportional valve; 3 PAM; 4 Angle transducer; 5 Mechanical platform; 6 Air compressor; 7 Pressure conditioner )

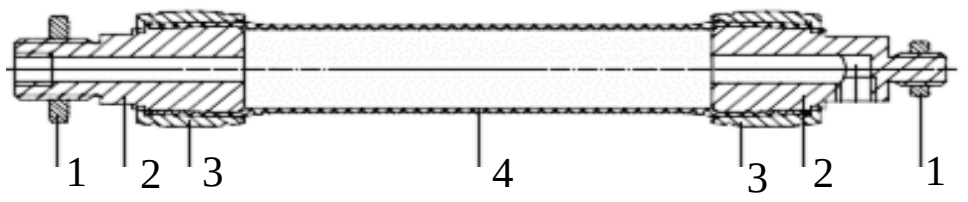

**Fig.2** The sectional view of PAM
(1.nut 2.flange 3.sleeve 4.membrane)

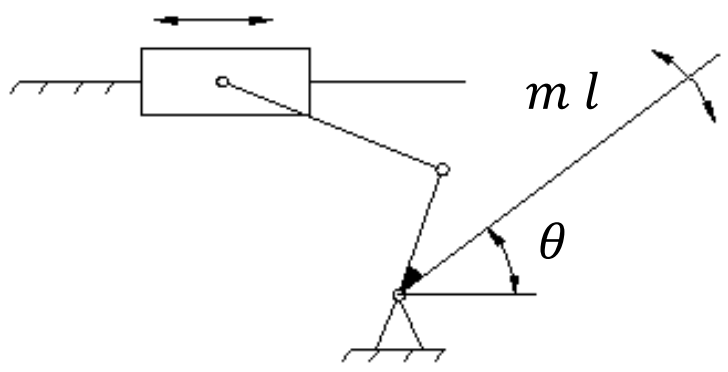

**Fig.3** Schematic of the mechanical pendulum platform

**Table 1** Structure parameters of mechanical platform

| $l$ / mm | $m$ / kg | $\theta$ / rad |
|---|---|---|
| 250 | 0.471 | 0< $\theta$<$\pi/2$ |

**2.2 Modelling the pneumatic system**

Considering the gas is ideal and isothermal and mass constant inside the PAM, the model between the pressure and volume of the isentropic and polytropic process is described as[12]:

$$p_1 V_1^{\kappa} = p_2 V_2^{\kappa} \tag{1}$$

where $p_i$ and $V_i$ denotes pressure and volume at the $i$th state, and $\kappa$ the polytropic index. Meanwhile, the relation between mass, pressure, temperature, and volume is given by:

$$m_g = \frac{pV}{RT} \tag{2}$$

where $m_g$ is the mass of gas inside the muscle, $T$ the gas temperature and $R$ the gas constant. Combined with (1), the dynamic model of the pressure can be expressed by the total differential form of (2) as:

$$\frac{dp}{dt} = \frac{\kappa}{V(s)}\left[RT\dot{m}_g - p\frac{dV}{ds}\dot{s}\right] \tag{3}$$

where $s$ is the contraction displacement of PAM and $\dot{m}_g$ is the mass flow rate which is determined by the structure of valve and can be identified by experiment. Set $\dot{m}_g = Q_m(u,p)$ which can be calculated as:

$$Q_m = \upsilon \cdot \eta \cdot p_s \cdot \psi \tag{4}$$

with

$$\psi = \sqrt{\frac{2\kappa}{RT(\kappa-1)}\left[\left(\frac{P}{P_S}\right)^{\frac{2}{\kappa}} - \left(\frac{P}{P_S}\right)^{\frac{\kappa+1}{\kappa}}\right]} \tag{5}$$

where $\upsilon$ is the input voltage of valve, $\eta$ the discharge coefficient, $p/p_s$ the ratio between the downstream pressure $p$ and the upstream pressure $p_s$ .

In this paper, the gas flow is controlled with an electronic proportional directional control valve in 5/3-way produced by the pneumatic manufacturer FESTO[13-14]. The setpoint value input as analogue voltage signal is from 0v to 10v and the standard nominal rated flow 350L/min. The signals is collected with the software of NI LABVIEW and hardware of NI cDAQ-9178 compact data acquisition platform with NI 9269 and NI 9205 modules.

In (1) and (2), coefficients $T, R,$ and $\kappa$ are defined by experience, and the discharge coefficient $\eta$ determined by fitting the relationship between theoretical and experimental port area, as shown in Fig 4. All coefficients are listed in Table 2.

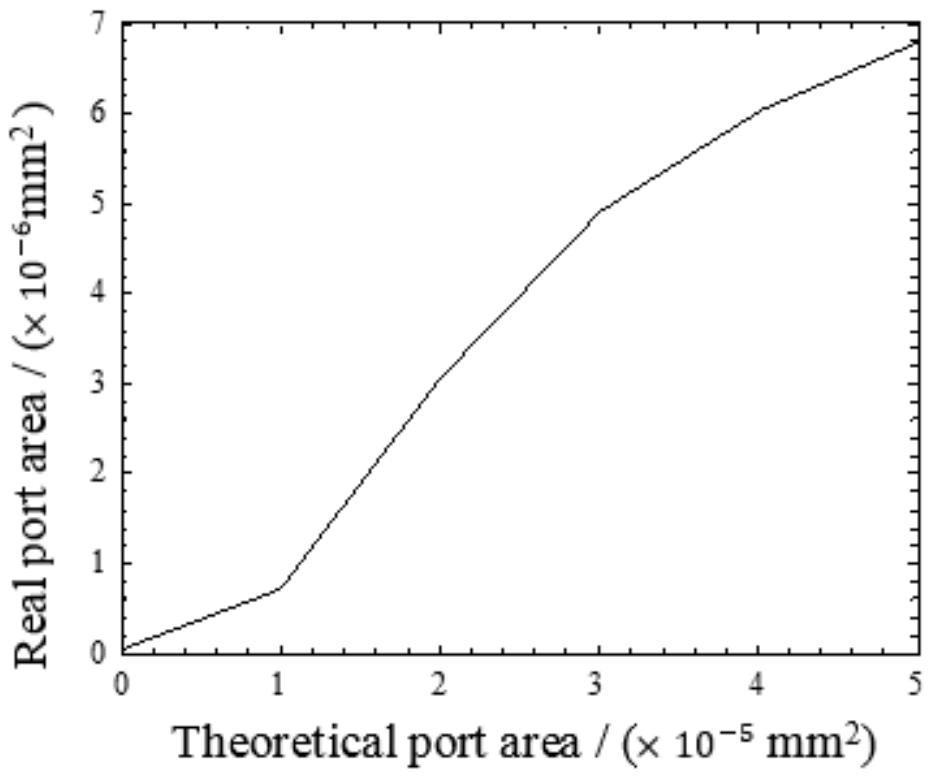

**Fig.4** The calibration of discharge coefficient

**Table 2** Coefficients of pneumatic circuit

| $T$ / K | $R$ /(J · kg$^{-1}$ · | $\kappa$ | $\eta$ |
|---|---|---|---|
| 293 | 287.1 | 1.4 | 0.14 |

**2.3 Modelling the dynamics system**

To represent the nonlinear relationship between the motion output and inner driving pressure input with linear model, firstly, it can be approximated by a parameter-varying model[15-16].

Set

$$x = [\theta, \dot{\theta}, \ddot{\theta}]^T \quad (6)$$

then

$$\begin{aligned} p = A(x) &\approx F_J(\ddot{\theta}) + F_C(\dot{\theta}) + F_K(\theta) \\ &= \overline{J}(\theta)\ddot{\theta} + \overline{C}(\theta)\dot{\theta} + \overline{K}(\theta)\theta + \sigma(x) \\ &= \overline{A}(\theta)x + \sigma(x) \end{aligned} \quad (7)$$

where

$$\sigma(x) = f(\theta^{(n)})x\,, \quad (n > 2) \quad (8)$$

where $p$ denotes the driving pressure, $\theta$ the angular position of the pendulum, $A$ and $\overline{A}$ the real nonlinear and approximated linear parameter-varying model, $F(\cdot)$ denotes the nonlinear submodel related to different order, $\overline{J}$, $\overline{C}$, and $\overline{K}$ the parameter-varying linear submodel, and $\sigma$ the high-order term.

Then, the transfer function of the linear model is expressed as

$$G(s) = \frac{1}{\underbrace{\overline{J}(\theta)s^2 + \overline{C}(\theta)s + \overline{K}(\theta)}_{part\ 1} + \underbrace{\delta(s^n)}_{part\ 2}}\,, (n \geq 3) \quad (9)$$

And the nonlinearity of the system is model into the effect of parameters deviation related to angular position, as $part\ 1$, accompanied with a high order part, as $part\ 2$.

To describe the $part\ 1$, the model set modeling(MEM) method is applied, by which the real model can be described with a set of possible submodels[17]. Based on it, a set of second order submodels are identified in corresponding sub-workspaces. In the modelling experiment, the range of input pressure is divide into 10 successive parts, and these signals of input and output are plotted in Fig 5. The parameters of characteristic polynomial of each model are listed in Table 3 and the mean value $\bar{z}$($\bar{z} = \sum_1^N z_i/N$, $z$=$j$, $c$, and $k$) and the maximum relative error $\widehat{e_r}$ ( $\widehat{e_r} = \max_{1 \leq i \leq N}(|z_i - \bar{z}|/\bar{z})$, $z$=$j$, $c$, and $k$) of them are calculated as well. The type of PAM used in the experiment is DMSP-20-200N-AM-CM produced by the pneumatic manufacturer FESTO. And the angle transducer and pressure sensor is produced by SAKAE and SMC manufacturer respectively.

For the variation of parameters is bounded and ignoring the influence of $part\ 2$, the $G(s)$ can be transformed into a nominal model accompanied with the parametric perturbations, expressed as[18]

$$G_L(s) = \frac{1}{j_m(1 + p_J\delta_J)s^2 + c_m(1 + p_c\delta_c)s + k_m(1 + p_k\delta_k)} \quad (10)$$

with

$$-1 \leq \delta_j, \delta_c, \delta_J \leq 1$$

where $j_m$, $c_m$, and $k_m$denotes the mean value error of each parameters, and $p_j$,$p_c$ and $p_k$ the maximum relative of them respectively.

**Table 3** parameters of characteristic polynomials

| | $j$ | $c$ | $k$ |
|---|---|---|---|
| Model 01 | 0.0023 | 0.0423 | 3.7428 |
| Model 02 | 0.0025 | 0.0394 | 3.4660 |
| Model 03 | 0.0023 | 0.0386 | 2.7861 |
| Model 04 | 0.0022 | 0.0348 | 2.3160 |
| Model 05 | 0.0022 | 0.0358 | 2.0670 |
| Model 06 | 0.0025 | 0.0378 | 1.9153 |
| Model 07 | 0.0022 | 0.0480 | 1.7545 |
| Model 08 | 0.0025 | 0.0550 | 1.6836 |
| Model 09 | 0.0029 | 0.0354 | 1.5652 |
| Model 10 | 0.0037 | 0.0339 | 1.3847 |
| $\bar{z}$ | 0.0025 | 0.0445 | 2.5638 |
| $\widehat{e_r}$ | 0.2535 | 0.2382 | 0.4599 |

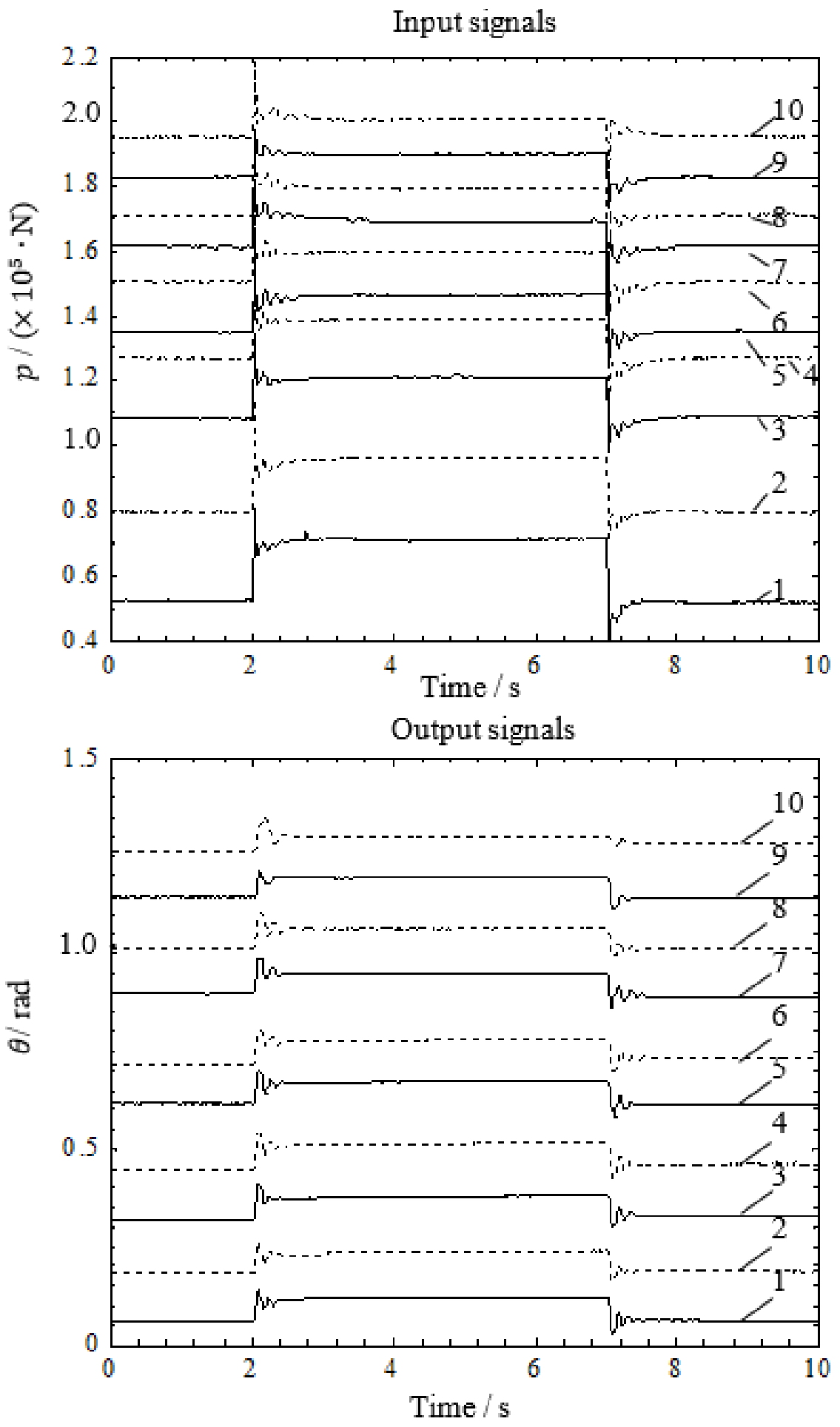

**Fig.5** Input / Output signals of Pressure / Angle
(the signals of input and output are one-to-one corresponding from the bottom to the top)

## 3 Controller design

Cascaded control strategy is effective when the relationship between the actual input and output of the system is complicated but can be simplified by described with cascaded submodels[19]. In this article, a cascaded control strategy is applied, of which the inner-loop is designed for driving pressure tracking inside the PAM and an outer-loop for angular position tracking of mechanical structure. The closed-loop system is described in Fig 6

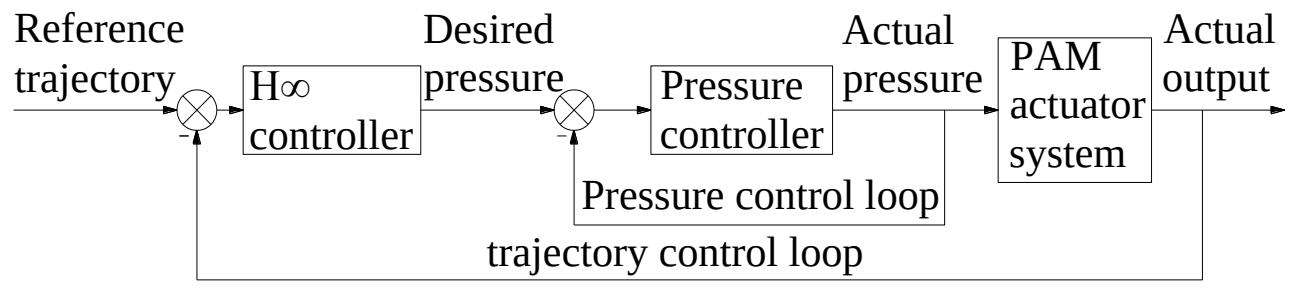

**Fig.6** Cascaded control loop of nonlinear system

### 3.1 Pressure control

Feedback linearization control strategy is broadly used for nonlinear system control, which can rearrange the system by introducing an external reference related to the final output indirectly. Based on it and combined with (3), (4) and (5), the derivative of pressure is chosen as the external reference signal of the inner loop for pressure tracking:

$$r = \dot{p} \tag{11}$$

where $r$ denotes the external reference input. To diminishing the steady error exponentially, the external input is written as:

$$r = \dot{p}_d + K_p(p_d - p) \tag{12}$$

where $K_p$ is the control parameter and $p_d$ the desired pressure. And the final input voltage of valve is given as:

$$v = \frac{[\dot{p}_d + K_p(p_d - p)] \cdot \sum_{i=0}^{5} b_i s^i + \gamma \cdot p \cdot \dot{s} \cdot \sum_{i=1}^{5} i b_i s^{i-1}}{\gamma \cdot \eta \cdot p_s \cdot \sqrt{\frac{2\kappa \cdot RT}{(\kappa-1)}\left[\left(\frac{P}{P_S}\right)^{\frac{2}{\kappa}} - \left(\frac{P}{P_S}\right)^{\frac{\kappa+1}{\kappa}}\right]}} \tag{13}$$

### 3.2 Trajectory tracking control

For suppress the effect of parametric perturbations, an $\mathcal{H}\infty$ controller is synthesize based on the loop-shaping design procedure (LSDP), and the principle diagram of it shown in Fig 7:

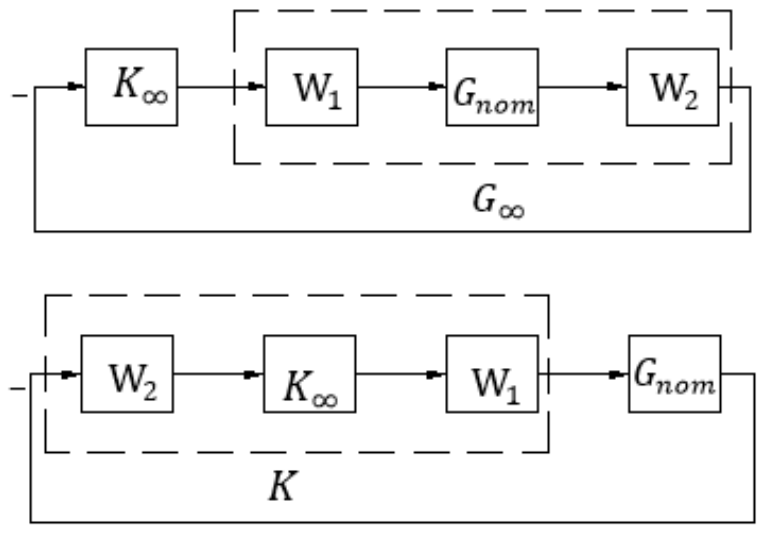

**Fig.7** Schematic block diagram of LSDP
(Upper: reshaping the open-loop; Lower: synthesizing the final controller)

W1 and W2 is the pre- and post-compensator respectively, which are used to reshape the frequency response of open-loop transfer function so that the closed-loop system can achieve the required performance in different frequencies. $\boldsymbol{G}_\infty$ is the shaped plant and formed as $\boldsymbol{G}_\infty = \boldsymbol{W_2 G W_1}$, $\boldsymbol{K}_\infty$ the $\mathcal{H}\infty$ controller to stabilize the reshaped plant robustly, and $\boldsymbol{K}_{final}$ the final feedback controller formed as:

$$\boldsymbol{K} = W_1\, \boldsymbol{K}_\infty W_2 \tag{14}$$

Consider the robust stabilization criterion[20]: Let $\boldsymbol{G}_{nom} = \widetilde{\boldsymbol{M}}^{-1}\widetilde{\boldsymbol{N}}$ be the nominal model and the perturbed system can expressed as

$$\boldsymbol{G}_\Delta = \left(\widetilde{\boldsymbol{M}} + \widetilde{\Delta}_{\boldsymbol{M}}\right)^{-1}\left(\widetilde{\boldsymbol{N}} + \widetilde{\Delta}_{\boldsymbol{N}}\right) \tag{15}$$

where $G_\Delta$ is the perturbed system. $\widetilde{\Delta}_{\boldsymbol{M}}$ and $\widetilde{\Delta}_{\boldsymbol{M}}$ are the perturbations of model, unknown but stable. $\widetilde{M}$, $\widetilde{N}$, $\widetilde{\Delta}_M$, $\widetilde{\Delta}_M \in \mathcal{RH}_\infty$ , and $\|[\widetilde{\Delta}_{\boldsymbol{M}} \quad \widetilde{\Delta}_{\boldsymbol{M}}]\|_\infty < \epsilon$ . And the perturbed system is robustly stable if and only if

$$\gamma = \left\| \begin{bmatrix} \boldsymbol{K} \\ \boldsymbol{I} \end{bmatrix} (\boldsymbol{I} - \boldsymbol{G}_{nom}\boldsymbol{K})^{-1} \widetilde{\boldsymbol{M}}^{-1} \right\|_\infty \leq \frac{1}{\epsilon} \tag{16}$$

where$K$ is the feedback controller mentioned before and $\epsilon$ $(\epsilon > 0)$ the stability margin.

Applying linear factional transformation (LFT), the three perturbed coefficients of (10) can be expressed as (17) and the block diagram of it is shown in Fig 8.

$$\begin{gathered} \begin{bmatrix} y_J \\ \ddot{\theta} \end{bmatrix} = \begin{bmatrix} -p_j & 1/j_m \\ -p_j & 1/j_m \end{bmatrix} \begin{bmatrix} u_j \\ u - v_c - v_k \end{bmatrix} \\ \begin{bmatrix} y_c \\ v_c \end{bmatrix} = \begin{bmatrix} 0 & c_m \\ -p_c & c_m \end{bmatrix} \begin{bmatrix} u_c \\ \dot{\theta} \end{bmatrix} \\ \begin{bmatrix} y_k \\ v_k \end{bmatrix} = \begin{bmatrix} 0 & k_m \\ -p_k & k_m \end{bmatrix} \begin{bmatrix} u_k \\ \theta \end{bmatrix} \\ u_J = y_J \delta_J, \;\; u_c = y_c \delta_c, \;\; u_k = y_k \delta_k \end{gathered} \tag{17}$$

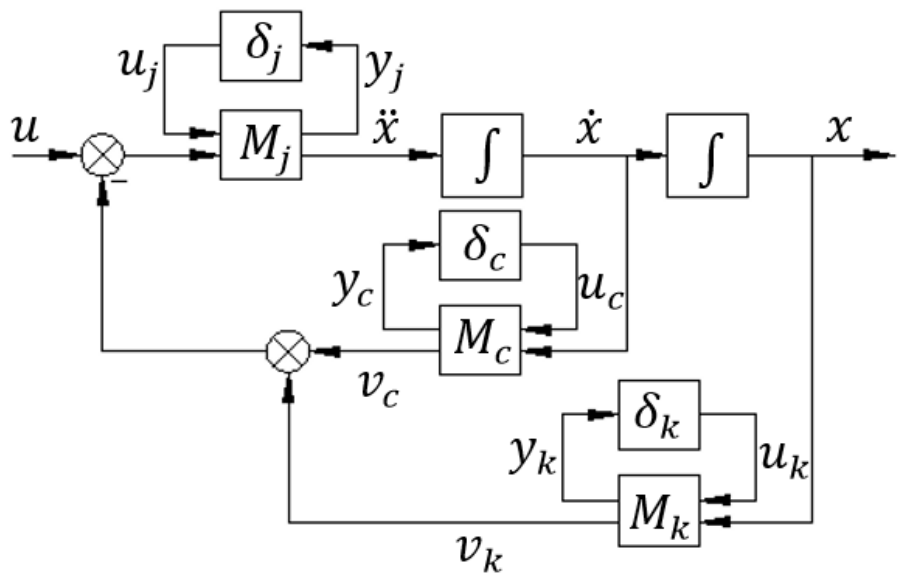

**Fig.8** Block diagram of model with parametric perturbations

where $y_J, y_c, y_k$ and $u_J, u_c, u_k$ denote the inputs and outputs of perturbation block $\delta_J$, $\delta_c$ , $\delta_k$, respectively.

Let

$$\begin{cases} x_1 = \theta, \quad x_2 = \dot{\theta} = \dot{x}_1, \quad y = x_1 \\ \dot{x}_2 = \ddot{\theta} = \ddot{x}_1 \end{cases}$$

The state-space equation governing the dynamics behavior of mechanical system are given by:

$$\begin{bmatrix}\dot{x}_1\\ \dot{x}_2\\ y_j\\ y_c\\ y_k\\ y\end{bmatrix} = \begin{bmatrix} 0 & 1 & 0 & 0 & 0 & 0\\ -\frac{k_m}{j_m} & -\frac{c_m}{j_m} & -\frac{c_m}{j_m} & -\frac{p_c}{j_m} & -\frac{p_k}{j_m} & \frac{1}{j_m}\\ -\frac{k_m}{j_m} & -\frac{c_m}{j_m} & -p_j & -\frac{p_c}{j_m} & -\frac{p_k}{j_m} & \frac{1}{j_m}\\ 0 & c_m & 0 & 0 & 0 & 0\\ k_m & 0 & 0 & 0 & 0 & 0\\ 1 & 0 & 0 & 0 & 0 & 0 \end{bmatrix} \begin{bmatrix} x_1\\ x_2\\ u_j\\ u_c\\ u_k\\ u\end{bmatrix} \quad (18)$$

$$\begin{bmatrix}u_j\\ u_c\\ u_k\end{bmatrix} = \Delta\cdot\begin{bmatrix}y_j\\ y_c\\ y_k\end{bmatrix} = \begin{bmatrix}\delta_j & 0 & 0\\ 0 & \delta_c & 0\\ 0 & 0 & \delta_k\end{bmatrix}\begin{bmatrix}y_j\\ y_c\\ y_k\end{bmatrix}$$

If set

$$\begin{cases}\widetilde{\boldsymbol{N}} = \boldsymbol{I}, \quad \widetilde{\Delta}_{\boldsymbol{N}} = \boldsymbol{0}\\ \widetilde{\boldsymbol{M}}^{-1} = \boldsymbol{G}_{nom}\end{cases} \quad (19)$$

then, the system can be reconstructed as shown in Fig 9

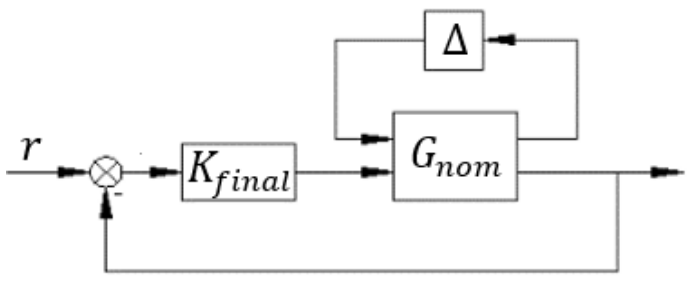

**Fig.9** Closed-loop system structure

and the criterion of robust stability is developed as:

$$\begin{aligned} &\left\|\begin{bmatrix}\boldsymbol{I}\\ \boldsymbol{K}_\infty\end{bmatrix}(\boldsymbol{I}-\boldsymbol{G}_\infty\boldsymbol{K}_\infty)^{-1}\widetilde{\boldsymbol{M}}_\infty{}^{-1}\right\|_\infty \\ &= \left\|\begin{bmatrix}I\\ \boldsymbol{K}_\infty\end{bmatrix}(\boldsymbol{I}-\boldsymbol{G}_s\boldsymbol{K}_\infty)^{-1}[\boldsymbol{I} \quad \boldsymbol{G}_\infty]\right\|_\infty \\ &= \left\|\begin{bmatrix}W_2 & \\ & W_1^{-1}\end{bmatrix}\begin{bmatrix}\boldsymbol{I}\\ \boldsymbol{K}\end{bmatrix}(\boldsymbol{I}+\boldsymbol{G}_{nom}\boldsymbol{K})^{-1}[\boldsymbol{I}\ \boldsymbol{G}_{nom}]\begin{bmatrix}W_2 & \\ & W_1\end{bmatrix}\right\|_\infty \\ &\le \epsilon^{-1} \end{aligned} \quad (20)$$

Ignoring the influence from high order part, as $part\ 2$, the post-compensator $W_2$ is set to 1 in the article. For the nominal model is only second order, the pre-compensator W1 is set to first order, expressed as:

$$W_1 = \frac{(1/M)s + w_0}{s + A} \quad (21)$$

To eliminate the steady-state error, increasing the gain of shaped plant over low frequency range is an effective way. Meanwhile, to improve the transient performance, a higher cross-over frequency is accepted, and yet exaggerates the influence of parameters varying with high frequencies. Thus a tradeoff between them must be faced. However, for the bandwidth of the controlled inner pneumatic circuit is found about 3Hz by sine sweep frequency experiment. For the bandwidth is lower than the natural frequency of the mechanical platform, the 3Hz is set as the cross-over frequency of the controlled system. To study the effect of the gain over high frequency-range on the transient performance, three different $W_1$s are selected and listed in Table 3. The bode diagrams of the three shaped plants(SP) are shown in Fig 10. For illustrating the transient performance of these controllers synthesized with different $W_1$ s, the step responses of these closed-loops with corresponding controllers are shown in Fig 11, where each curve of a subfigure corresponds to a model expressed as (10) with certain perturbed parameters. For generality, the mean, low and high bound of each parameters are considered, and $3^3$ curves are depicted in each subplot.

Comparing the three shaped systems, the three cases have the same steady-state error and rise time. However, with the gain lower over high-frequency range, the output is smoother, which can be confirmed by comparing the angular acceleration as shown in Fig 12. During the rise phase, the rate of velocity change of case 3 is smaller than the rest. Thus, the controller $K_3$ synthesized by $W_{1_3}$ is chosen. And the $1/\gamma$ is 0.7154 satisfying the requirement of inverse stability margin $\epsilon_{max}$. The singular value of each components in the left of (21) is also calculated and shown in Fig 13, where the sensitivity of closed-loop over high-frequency range and the sensitivity complementary over low-frequency are all smaller than 1, and the theoretical bandwidth of the closed-loop system is about 1.2 rad/s.

**Table 3 Parameters of W1**

| | $M$ | $w_0$ | $A$ |
|---|---|---|---|
| $W_{1_1}$ | 0.4 | 3 | 0 |
| $W_{1_2}$ | 1 | 3 | 0 |
| $W_{1_3}$ | 8 | 3 | 0 |

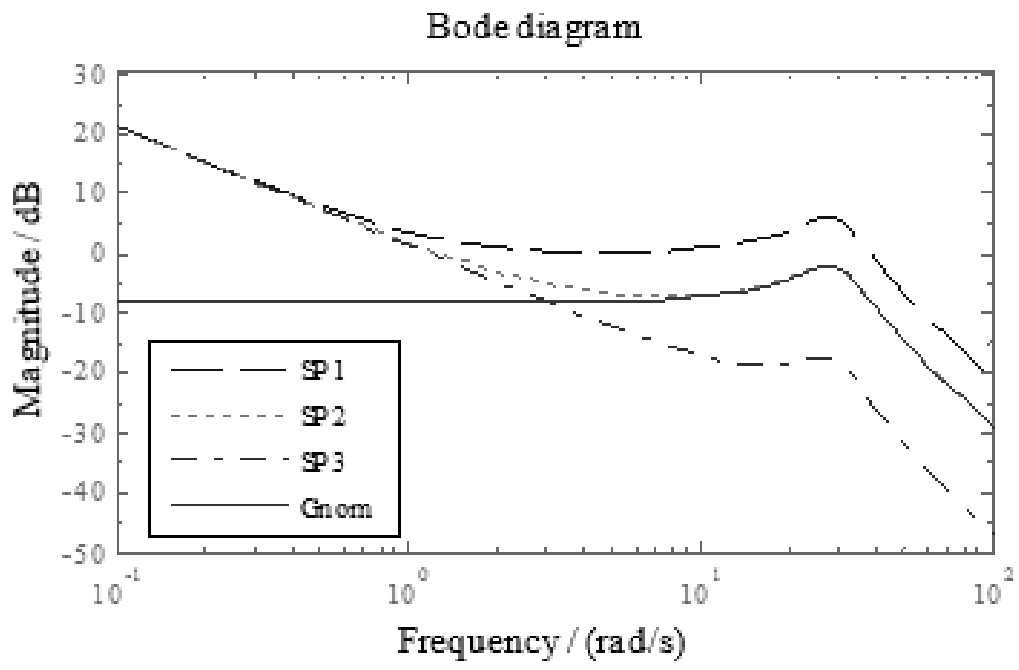

**Fig.10** Bode diagrams of plants shaped with W1

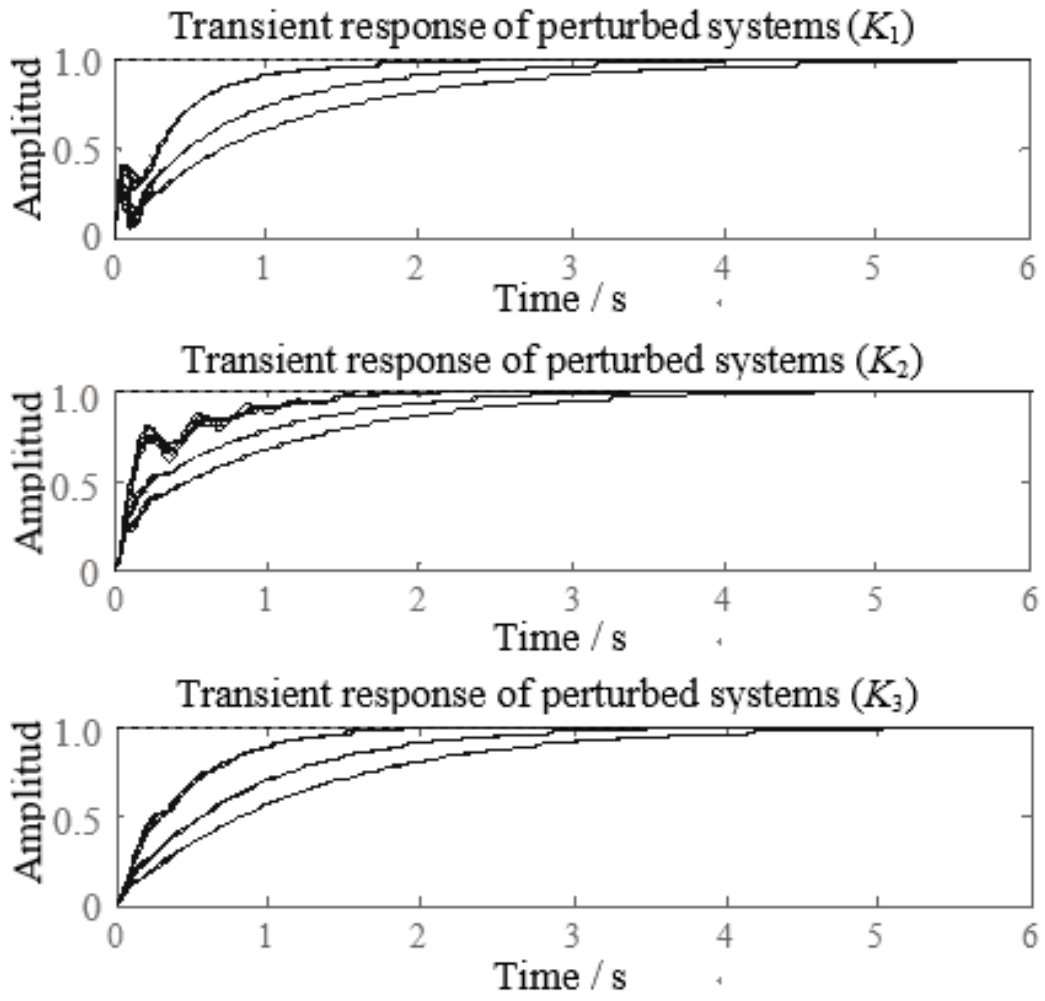

**Fig.11** Transient response of perturbed systems (top: controlled by $K_1$; Medium: controlled by $K_2$; bottom: controlled by $K_3$)

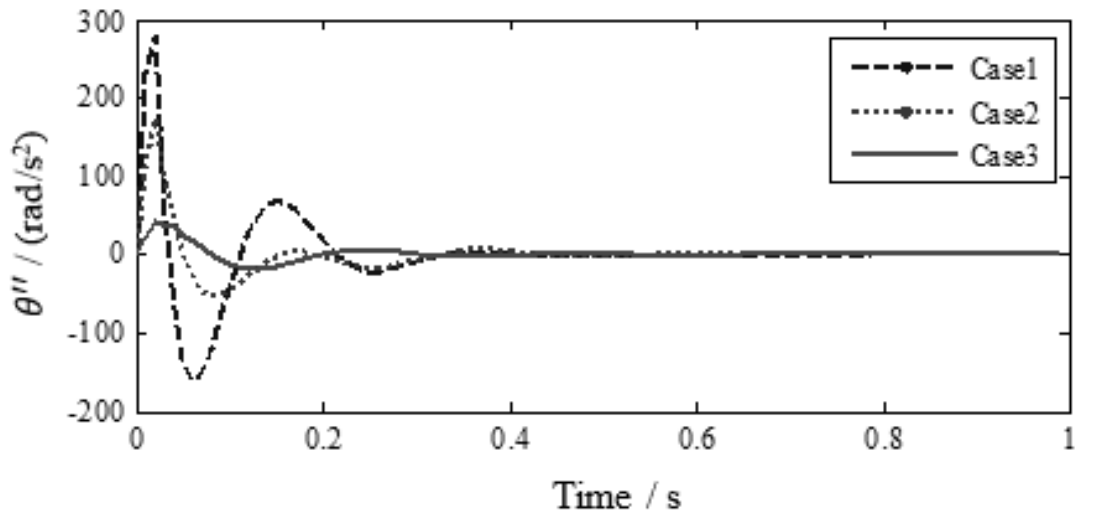

**Fig.12** Angular acceleration

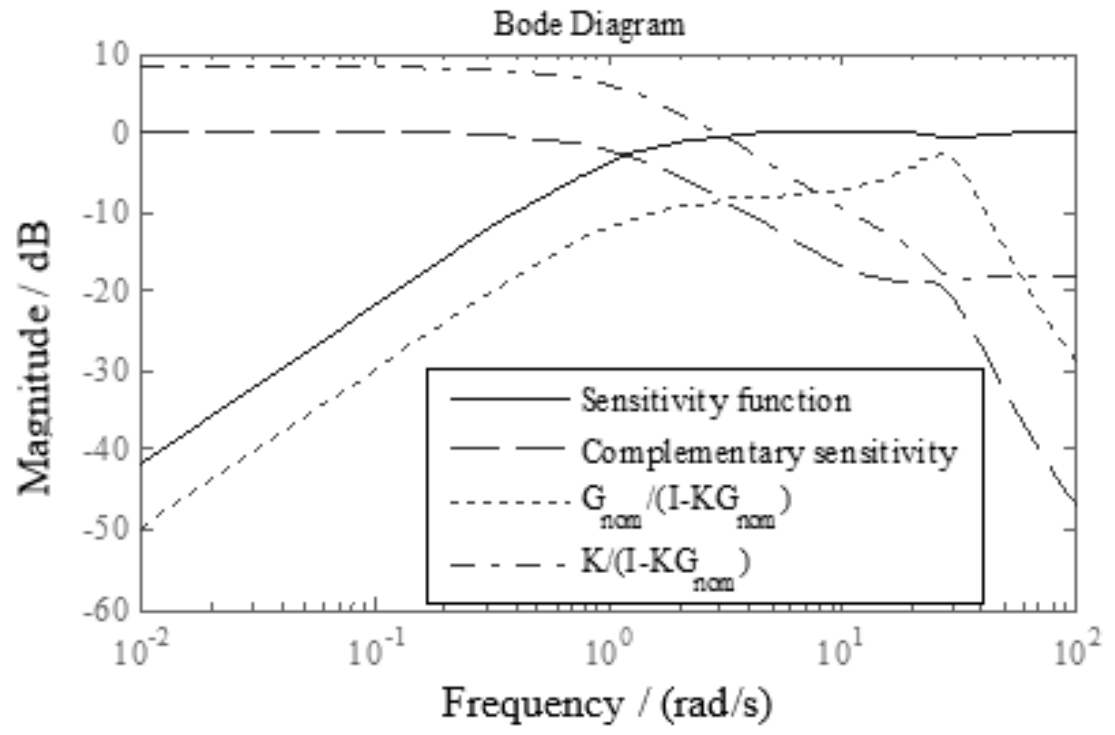

**Fig.13** Singular values of functions in robustly stable criterion

# 4 Experiment and Discussion

To demonstrate the performance of this closed-loop system, a signal composed with 10 different desired steady-states ranging from 0.1rad to 1.4rad is designed, as shown in Fig 14. For observing the error for velocity tracking, cosine curve is used to connect every two neighboring steady-states, which makes the desired trajectory is differentiable during the whole tracking process. The performance of angular position tracking and velocity tracking are shown in Fig 14, 15, and 16. The Fig 15 only shows the relative error of steady-states for trajectory tracking.

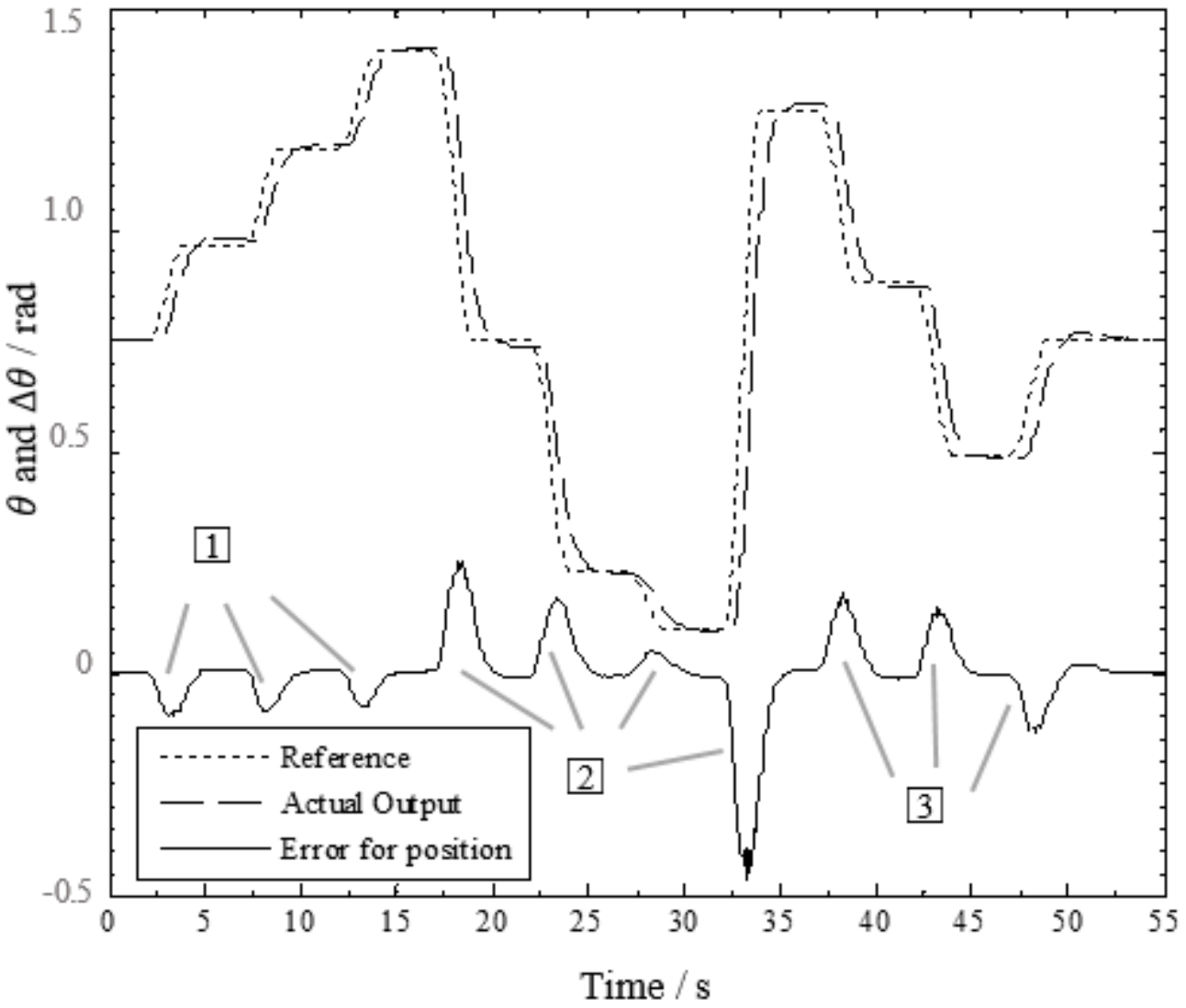

**Fig.14** Trajectories tracking of angular position
$(Error = Actual\ output_i - reference_i|)$

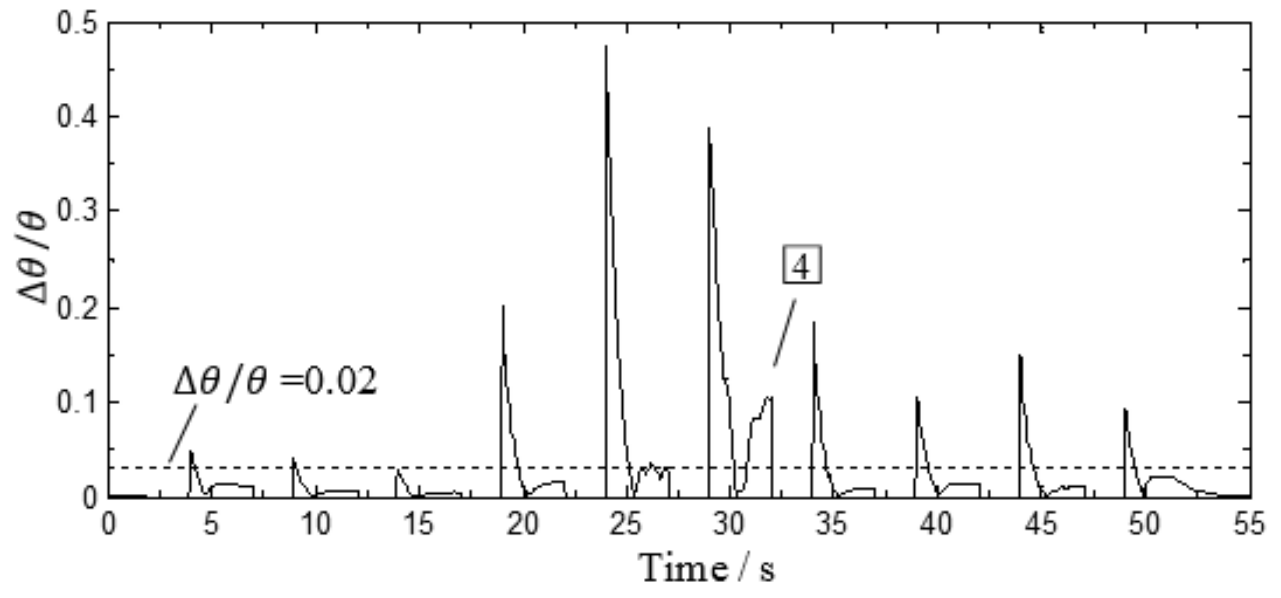

**Fig.15** Relative error for position of steady-states

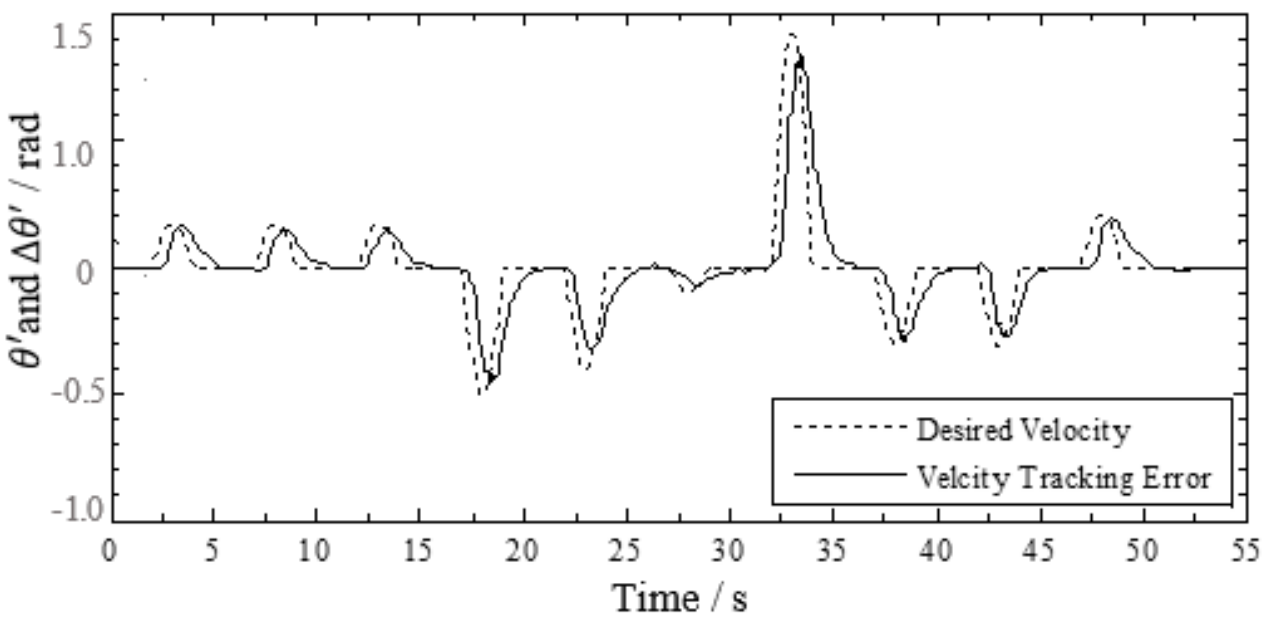

**Fig.16** Trajectory tracking of angular velocity

Except for 4⃞, the maximum of relative error during steady-state is smaller than 2% as shown in Fig 15, which shows the good robustness to the effect of parameter variation over low-frequency range. And, the small gain over high-frequency range results good smoothness and non-overshoot. However, for the transient performance sacrificed, a time delay exists between the desired and actual trajectory, which is the main source of tracking error for position and velocity. It leads to the maximums relative tracking errors for angular position appear when the sign of velocity tracking error changes as shown in Fig 14 and 16. The maximum of tracking error for trajectory is -0.46 during the interval from 0.1rad to 1.27rad for climb and 0.25 during the interval from 0.75rad to 0.23rad for descent

Meanwhile, some details for transient performance can be found. although there is a positive relation between the maximum of tracking error and the stroke between every two neighboring steady-states as shown by 2⃞ and 3⃞ in Fig 14, the relative errors decreases and velocity error stable when the position of these strokes rising, as shown by 1⃞, which indicates the transient performance is robustly stable when the real model varying. The maximum of tracking error for velocity during acceleration and deceleration is 1.458rad/s and 1.261rad/s respectively. After that these errors converge to zero exponentially, as shown in Fig 15 and 16.

For the leakage of the electronic proportional is unavoidable, the performance is deteriorated when the inner driving pressure low. And, the effect of it can be counteracted by applying symmetrical structure and using two PAMs in an agonist-antagonist way.

# 5 Conclusion

1) For simulating the motion mechanism of elbow musculoskeletal system, an unsymmetrical bionic joint actuated with a single PAM is built, and a back-and-forth movement of forearm in pitching plane is realized.

2) Based on the robust modeling theory, the nonlinear relationship between the inner driving pressure of PAM and the motion output is model as a second order linear time-invariant model accompanied with parametric perturbations, by which a modeling for the nonlinear system is realized

3) A cascaded controller, constructed with an outer $\mathcal{H}\infty$ controller synthesized by LSDP to reduce the effect of parametric perturbations and an inner controller based on the feedback linearization theory for the driving pressure control, is designed. And a trajectory tracking of the bionic joint is realized with it.

## Reference

[1] Caldwell D G, Medrano-Cerda G A, Goodwin M. Control of pneumatic muscle actuators[J]. IEEE Transactions on Control Systems, February 1995, 15: 40-48.

[2] Hildebrandt A, Sawodny O, Neumann R, Hartmann A. A flatness based design for tracking control of pneumatic muscle actuators[C]. 2002 IEEE International Conference on Control, Automation, Robotics and Vision, 2002, 3: 1156-1161.

[3] Schindele D, Aschemann H. Comparison of Cascaded Backstepping Control Approaches with Hysteresis Compensation for a Linear Axis with Pneumatic Muscles[C]. 2013 IFAC Symposium on Nonlinear Control Systems, 2013, 9(1): 773-778.

[4] Zhu Xiao-cong, Tao Guo-liang, Yao Bin, Cao Jian. Adaptive robust posture control of a parallel manipulator driven by pneumatic muscle[J]. Automatica, September 2008, 44(9):2248-2257.

[5] Schindele D, Aschemann H. Trajectory tracking of a pneumatically driven parallel robot using higher-order SMC[C]. 2010 IEEE International Conference on Methods and Models in Automation and Robotics, 2010: 387-392.

[6] Aschemann H, Schindele D. Comparison of model-based approaches to the compensation of hysteresis in the force characteristic of pneumatic muscles[J]. IEEE Transactions on Industrial Electronics, 2014, 61(7): 3620-3629.

[7] Chikh L, Poignet P, Pierrot F, Michelin M. A predictive robust cascade position-torque control strategy for Pneumatic Artificial Muscles[C]. 2010 IEEE American Control Conference, 2010: 6022-6029.

[8] Lin Liu-Hsu, Yen Jia-Yush, Wang Fu-Cheng. Robust control for a pneumatic muscle actuator system[J]. Transactions of the Canadian Society for Mechanical Engineering, 2013, 37:581-590.

[9] Slotine J, Li Wei-ping, Applied nonlinear control[M]. Englewood Cliffs. NJ: Prentice-Hall, 1991:14.

[10] Festo AG & Co. KG, Documentation Fluidic Muscle DMSP 2013 [EB/OL], http://www.festo.com.cn/cat/en-cn_cn/data/doc_engb/PDF/EN/DMSP-MAS_EN.PDF.

[11] Shin D, Seitz F, Khatib O, et al. Analysis of torque capacities in hybrid actuation for human-friendly robot design[C]. IEEE International Conference on Robotics and Automation (ICRA), 2010: 799-804.

[12] Beater P, Pneumatic drives[M]. Berlin Heidelberg: Springer-Verlag, 2007:28-32.

[13] Festo AG & Co. KG, Documentation for Proportional directional control valves MPYE 2013[EB/OL], http://www.festo.com.cn/cat/en-cn_cn/data/doc_engb/PDF/EN/MPYE_EN.PDF.

[14] Krichel S V, Sawodny O, Hildebrandt A. Tracking control of a pneumatic muscle actuator using one servovalve[C]. IEEE American Control Conference (ACC), 2010: 4385-4390.

[15] Mar á T R, Banks S P. Linear, Time-varying Approximations to Nonlinear Dynamical Systems[J]. Control and Information Sciences, Berlin Heidelberg, Springer-Verlag, 2010,411:11-28

[16] Wang Xin, Li Man-tian, Guo Wei, P. Wang, and L. Sun. Development of an antagonistic bionic joint controller for a musculoskeletal quadruped[C]. 2013 IEEE/RSJ International Conference on Intelligent Robots and Systems, 2013: 4466-4471.

[17] Reinelt W, Garulli A, Ljung L. Comparing different approaches to model error modeling in robust identification[J]. Automatica. May 2002, 38(5):787-803.

[18] Gu Da-wei, Petkov P H, Konstantinov M, Robust Control Design with MATLAB[M]. London: Springer-Verlag, 2005:13-67.

[19] Hildebrandt A, Sawodny O, Neumann R, and Hartmann A. A cascaded tracking control concept for pneumatic muscle actuator[C]. 2003 European Control Conference, 2003: CD-ROM.

[20] Zhou Ke-ming, Doyle J C. Essentials of robust control[M]. Upper Saddle River, NJ: Prentice hall, 1998:226-235.